\documentclass{ws-ijmpe}
\usepackage[super,compress]{cite}
\begin{document}

\markboth{Authors' Names}{Instructions for typing manuscripts (paper's title)}

\catchline{}{}{}{}{}

\title{Nucleon properties inside compressed nuclear matter.}

\author{Jacek Ro\.zynek\footnote{pernament adress}}

\address{National Centre for Nuclear Research, Ho\.za 69, 00-681 Warsaw, Poland\\
first\_author@university.edu}
\maketitle

\begin{history}
\received{Day Month Year}
\revised{Day Month Year}
\end{history}

\begin{abstract}
 In this work we show the modifications of nucleon mass and
nucleon radius with the help of the extended Relativistic Mean
Field (RMF) model. We argue that even small departures above nuclear equilibrium density
with constant nucleon mass require an energy transfer from the repulsive mean field to the quarks forming nucleon massive bags in Nuclear Matter (NM), together with the
decrease in the nucleon volume. The transfer, which is proportional to pressure and absent in a standard RMF approach, provides good values for nuclear compressibility, symmetry energy and its slope. Different courses of the Equation of State (EOS), which depend
on the energy transfer, are considered.
\end{abstract}


\ccode {21.65.+f,24.85.+p}

\section{Introduction}
\noindent One assumption is common for most theoretical descriptions of NM
or finite nuclei: nucleons are treated as point particles with a bare mass $M_N$. The
kinematical description of nuclear reactions, model calculations of the single
particle (s.p.) spectrum and binding effects, indicate that the nucleon mass remains
unchanged inside the saturated (no pressure) nuclear medium. This, of course, does not
apply to an effective self-consistent nucleon mass $M^*_N$ in the RMF approach \cite{Ring,ms2}, which
contains contributions from the scalar field or to another ``effective" nucleon mass
used in the non-relativistic approach, which contains in addition part of the
s.p. interaction. The bare nucleon mass $M_N$ is the result of the strong interaction between
the almost massless quarks.  The nuclear deeply inelastic scattering (EMC effect) \cite{Jaffe,Fran} and nuclear Drell-Yan
experiments \cite{drell}, which measure the sea quark enhancement, can be described
\cite{jacek} with a small (1\%) admixture of nuclear pions and an unchanged nucleon mass $M_N$.
Thus, the deep inelastic phenomenology confirms [2 - 6] that the change in
the nucleon invariant bare mass in comparison to the value in vacuum is negligible at
the saturation density, although the scalar and vector mean fields are strong
\cite{wa}. Therefore we assume that a nucleon mass in nuclear medium $M_{pr}=M_N$. However, in a compressed medium the model assumption of point like nucleons is
difficult to accept because an additional work $W_N=p_HV_N$ of nuclear pressure $p_H$, which allows a
finite space $V_N$ for a nucleon ``bag", must be compensated for \cite{ja};
either by the energy of the nucleon constituents - quarks or by the meson field (or
both scenarios together). Existing work \cite{exvol,Costa,Costa1,benic,Ste,Typel} on finite volume considers
only fixed size and fixed mass nucleons without energy transfer. Chiral symmetry
restoration which can change additionally the  nucleon mass in the critical region of the phase transition  will be investigated in our model
in the future.

\noindent This analysis will involve functional corrections
to nucleon energies dependent on density with physical parameters - pressure,
nucleon radius $R$ and nucleon mass $M_{pr}$ in NM. Other modifications connected
with the finite volume of nucleons, like correlations of their volumes, will be
neglected.
For nucleon degrees of freedom we have the thermodynamical Hugenholz-van Hove\cite{HvH} (HvH) relation, which connects the chemical potential $\mu_N$  or
nucleon Fermi energy $E_F$,
with the nuclear single particle (s.p.) energy $\varepsilon_N=E_A/A$, pressure  $p\doteq\partial E_A/\partial
V$ and density $\varrho\doteq A/V=\gamma {P_F}^3/6\pi$ fm$^{-3}$:
\begin{eqnarray}
\!E_F\doteq\mu_N=\varepsilon_N+p/\varrho \label{Fermi1}
\end{eqnarray}
where $\gamma$ is a level degeneracy and $P_F$ is a Fermi momentum of point-like nucleons.

When the nucleons are extended objects of quarks the total nuclear energy $E_A/A$ and the nuclear single particle (s.p.) $\varepsilon^{q}_N=E_A/A$ are denoted by the additional index $q$.
The quarks  occupy a finite nucleon
volume $V_{N}$, therefore the available space for nucleon motion is
$V_{-}=V-AV_N$ smaller. Assuming the same meson exchange forces, which produce the correct value of the binding energy  for zero pressure,
the ``extended" nucleons interact in the smaller volume $V_{-}$, which causes effectively a bigger pressure $p_H$ above the equilibrium:
\begin{eqnarray}
\hspace{-2cm}p_H(\varrho)\doteq\left(\partial E_A/\partial
V_{-}\right)\simeq\hspace{1mm} \left(\partial
E_A/\partial V\right)(V/V_-)= p(\varrho)/(1-\varrho V_N(\varrho))
^{perfect}_{gas\hspace{0.5mm}appr.}
\label{press}
\end{eqnarray}
In the close-packing limit $\varrho \rightarrow 1/V_N$ and $p_H\rightarrow \infty$.
\section{A bag model in a nuclear medium}
 Describing nucleons as bags, pressure will influence their surfaces
\cite{Koch,Hua,Hua1,bag,bag1,Kap}.  In the bag model where the nucleon in the lowest state of three
quarks is a sphere of volume $V_{N}$ and its mass $M_N$ \cite{MIT,MIT1} is
a function of the radius $R_0$ with phenomenological constants - $\omega_0$, $Z_0$
\cite{Hua,Hua1} and a bag ``constant" $B(\varrho)$ which plays a role of a negative part of pressure inside a bag:
\begin{eqnarray}
M_N\!(R_0)\!\!&=&\frac{3\omega_0-Z_0}{R_0}+\frac{4\pi}{3}B(\varrho\!=\!0)R_0^3\propto~\!1/R_0.
\label{bag}
\end{eqnarray}
The following condition for the quark pressure $p_B$ inside a bag in equilibrium $(p_H=0)$,
\begin{eqnarray}
p_B=-\left(\partial M_N/\partial V_{N}\right)_{surface} =0
\label{pressured2}
\end{eqnarray}
measured on the surface, gives the relation
between $R_0$ and $B$, used at the end of (\ref{bag}).
In a compressed medium, the pressure generated by free quarks inside the bag \cite{MIT,MIT1}
is balanced at the bag surface not only by the intrinsic confining ``pressure"
$B(\varrho)$ but also by the nuclear pressure $p_H$; generated e.g. by elastic collisions
with other hadron \cite{Koch,Kap} bags, also derived in the QMC model in a medium
\cite{Hua}. When we assume that in a medium, internal parton pressure $p_B$ (\ref{pressured2}) inside the bag
is equal (cf. \cite{Hua,Hua1}) on the bag surface to the nuclear pressure $p_H$:
\begin{eqnarray}
p_H\!=p_B~\rightarrow ~R(\varrho)\!&=&\!\left[\frac{3\omega_0-Z_0}{4\pi (B(\varrho)+p_H(\varrho))}\right]^{1/4}
\label{pressure}
\end{eqnarray}
then we obtain the new, density dependent, bag radius $R$ (\ref{pressure}), depending from a sum $(B\!+\!p_H)$.
Thus, the pressure $p_H(\varrho)$ between the hadrons acts on the bag surface similarly
to the bag ``constant" $B(\varrho)$. A mass $M_{pr}(\rho)$ in nuclear medium
 can be obtained from (\ref{bag},\ref{pressure}):
\begin{eqnarray}
\hspace{-1mm}M_{pr}(\varrho)\!=\frac{3\omega_0-Z_0}{R{(\varrho)}}+
\frac{4\pi}{3}B(\varrho)R^3(\varrho)\!=\!M_N\frac{R_0}{R(\varrho)}\!-
\!p_q(\varrho)V_{N}(\varrho)\!=\!M_N\frac{R_0}{R(\varrho)}\!-\!W_N.
\label{massbag}
\end{eqnarray}
The volume energy $W_N(\varrho)=p_B V_N(\varrho)=p_H V_N(\varrho)$ is equal to the  work  which allows the creation of a bag volume in a compress nuclear medium.
Thus, the balance (\ref{pressure}) of nuclear pressure $p_H$ and quark pressure $p_B$  on the
bag surface determines the basic relationship (\ref{massbag}) between the changes of nucleon mass ($M_N\rightarrow M_{pr}$) and radius $(R_0\rightarrow R$) with the pressure $p_H$. This pressure, squeezing nucleon bags, induces the energy transfer from meson field
to quarks inside nucleons.
(Please note, that the equation (\ref{massbag}) defines quark chemical potential (enthalpy) $H_q\doteq M_{pr}(\varrho)+p_H V_N(\varrho)= M_{pr}(\varrho_0)R(\varrho_0)/R(\varrho)$).
Finally, for the extended nucleons with the density dependent mass $M_{pr}(\varrho)$ (\ref{massbag}) the HvH relation (\ref{Fermi1}) for the nucleon chemical potential takes the form:
\begin{eqnarray}
\!E^q_F\doteq\mu_N^q=\varepsilon^q_N\!+\!p_H V_{-}/A+p_B V_{N}
=\varepsilon^q_N\!+p_H/\varrho
\label{enthnuct}
\end{eqnarray}
\begin{figure}\hspace{.2cm}
\includegraphics[height=5.cm,width=6.3cm]{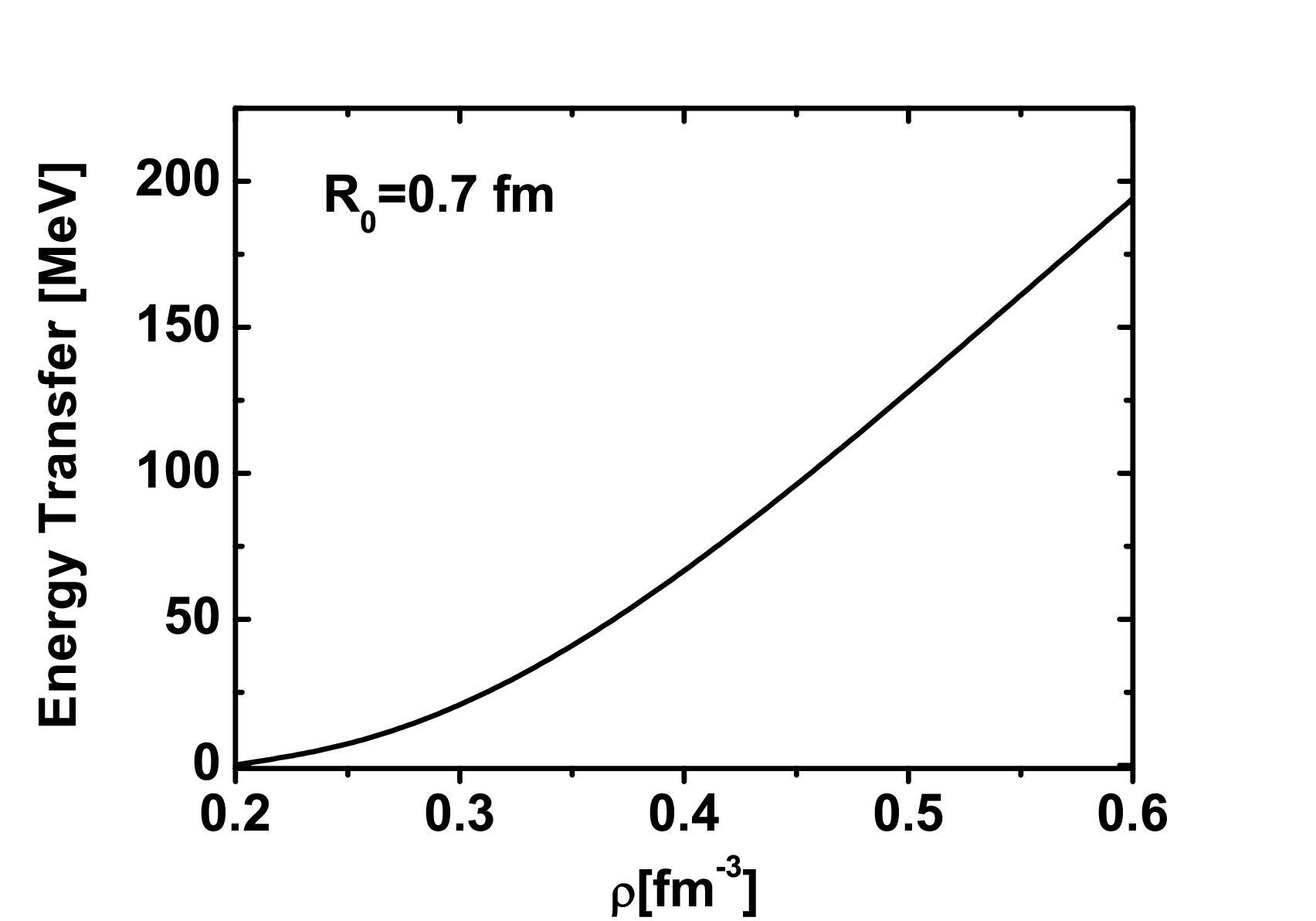} \caption{Scenario (\textbf{A}). Left, the energy transfer $\Delta E\!=\!W_N$ as a function of the NM density for an initial
nucleon radius $R_0=R(\varrho_0)=0.7$ fm and the const. mass $M_{pr}=M_N$.
Right, the pressure dependent radius $R$ for two initial
values of the $R_0$.} \label{bagmass}
\end{figure}
\begin{figure}
\vspace{-8.1cm}
\hspace*{6.2cm}\includegraphics[height=6.cm,width=6.3cm]{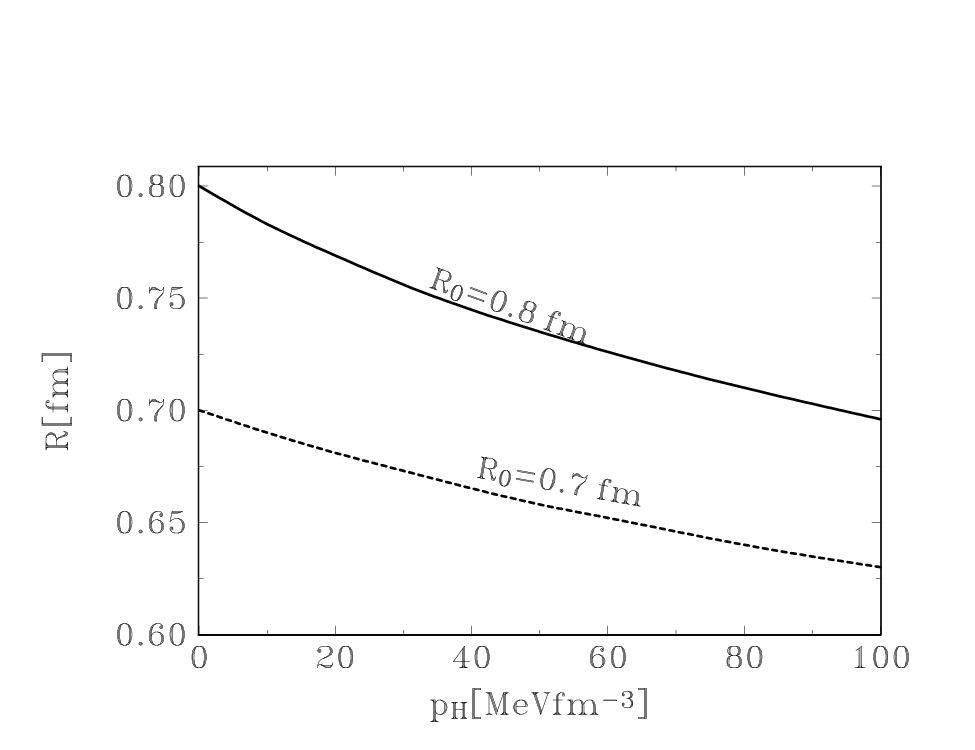}
\label{radius}
\vspace{1.2cm}
\end{figure}
\section{EoS with an energy transfer}
\noindent \noindent How does an energy transfer between quarks and the repulsive (or attractive) nuclear medium, influence the EoS? The complete answer will involve a complicated calculation. However, we have found a good estimate, by comparing \footnote{Our previous comparison \cite{ja} did not consider the energy transfer in case (A).} two extreme scenarios \textbf{(A)} and \textbf{(B)}.
In the scenario \textbf{(A)} the nucleon mass $M_{pr}(\varrho)=M_N$ is a constant, independent of
density. Therefore, the quarks inside the bag need
the additional (\ref{massbag}) energy transfer $\Delta E=p_H V_N(\varrho)$ to keep the constant mass and the bag
volume in the compressed medium. To increase the quark energy
from $M_N$ to $M_N+\Delta E$ respectively, a bag in NM should be compressed to a smaller radius
$R$ (right panel in Fig. \ref{bagmass}) (\ref{massbag}) by the repulsive interaction with the remaining nucleons. The accompanying energy transfer  above the equilibrium density $\varrho_0$, shown in left panel of Fig. \ref{bagmass}, will provide the volume energies $W_N$ inside the bags.
Finally, the s.p. energy $\varepsilon^q_N(\varrho)=E_A/A$, the Fermi energy $E^q_F(\varrho)$ and the radius $R(\varrho)$ (\ref{massbag}) can be written as \footnote{We extend  in Egs.(\ref{eq}-\ref{eq2}) the linear scalar-vector version of RMF with $\rho$ meson contributions to the symmetry energy \cite{wa,Kubis,Mo,Bielich,Bielich1}. Parameters  $(g_v/m_v)^2$,  $(g_s/m_s)^2$ are fitted at saturation point ($\varrho_0\approx 0.193$ fm$^{-3}$, $\varepsilon^q_N=15.6$ MeV); $g_v$($g_s$) is a vector(scalar) coupling const., $m_{v,s}$ - meson masses.}:
\begin{eqnarray}
\hspace{-5mm}\varepsilon^q_N(\varrho)&=&\frac{g_v^2}{2m_v^2}\varrho
\!+\!\frac{m_s^2}{g_s^2\varrho}(M_{pr}\!-\!M_{pr}^*)^2\!\!+
\!\frac{\gamma}{\varrho}\!\!\int_0^{P_F}\!\frac{d^3\!\mbox{\emph{\textbf{P}}}\!_N}{(2\pi)^3}\sqrt{\mbox{\emph{\textbf{P}}}_N^2\!+\!{M_{pr}^{*2}}}
-\Delta E(\varrho)  \label{eq}
\\ \hspace{-11mm}M^*_{pr}\!\!&=&\!M_{pr}-\frac{\gamma g_s^2}{2m_s^2}
\int_0^{P_F}\!\frac{d^3\!\mbox{\emph{\textbf{P}}}\!_N}{(2\pi)^3}
\frac{M^*_{pr}}{\sqrt{\mbox{\emph{\textbf{P}}}_N^2\!+\!M^{*2}_{pr}}}.
\\ \hspace{-15mm}E^q_F(\varrho)&=&\frac{g_v^2}{m_v^2}\varrho+
\sqrt{P_F^2\!+\!{M_{pr}^{*2}}}
-\Delta E(\varrho) \label{eq1}\\ \hspace{-19mm}
\vspace{0mm} R^{}_0/R(\varrho)&=&1+\Delta E(\varrho)/M_{pr}(\varrho) \hspace{6mm} \textit{where} \hspace{6mm} \Delta E(\varrho)=p_H V_N= \frac{\varrho^2{\varepsilon^{q}_N}^{'}\hspace{-1mm}(\varrho)\hspace{1mm}V_N(\varrho)}{(1-\varrho V_N(\varrho))}
\label{eq2}
\end{eqnarray}
For negative pressure $p_H$
 the nucleon bag increases its radius (\ref{massbag}), so the energy is transfer in opposite direction - from  bags to the meson field.
Summarizing, our model in scenario (\textbf{A}) consists of four self-consistent
equations (\ref{eq}-\ref{eq2}).

In order to show a thermodynamical consistency let us express the chemical potential $\mu_N^q$ (\ref{enthnuct},\ref{eq}) by the uniform pressure $p(\varrho)=\varrho^2 (\varepsilon^{q}_N)^{'}(\varrho)\!$ (\ref{press}):
\begin{eqnarray}
\hspace{-4mm}\mu_N^q=\varepsilon^q_N\!+p_H/\varrho=\varepsilon_N\!-p_H V_{N}+\!p_H V/A=\varepsilon_N\!+p_H V_{-}/A
=\varepsilon_N\!+(pV)/A
\label{chemical}
\end{eqnarray}
The formal dependence  from pressure are the same for $\mu_N^q(p)$ and $\mu_N(p)$ (\ref{Fermi1},\ref{chemical}). However, s.p. energies $\varepsilon^q_N(\varrho)$ of nucleons with finite volumes are smaller then s.p. energies $\varepsilon_N(\varrho)$ of point-like nucleons, by the volume energy $W_N=p_HV_N=\Delta E$ (\ref{eq}). Therefore, the total energy $E_A$, resulting pressure $p=\partial E_A/\partial V$ and the chemical potential (\ref{chemical},\ref{eq1}) are also smaller ($\mu_N^q<\mu_N$) with excluded volume corrections.
 In the uniform system of NM  the grand potential $\Omega=-pV$ given by a relation (\ref{chemical})
with $d\Omega=-pdV-Ad\mu_N^q$, satisfies the following thermodynamical relation for the average number of particles $A$,
\begin{eqnarray}
A=-\left[\frac{\partial\Omega}{\partial
\mu_N^q} \right]_V=\left[\frac{\partial (pV)}{\partial
\mu_N^q} \right]_V
\label{}
\end{eqnarray}
 which proofs the thermodynamical consistency of our model.

The energy transfer $\Delta E$ was not taken into
account in our previous findings in scenario  \textbf{(A)} presented in \cite{ja}. Let us compare these new results obtained in scenario
\textbf{(A)} with the results obtained without energy transfer, $\Delta E(\varrho)=0$ ,
when the nucleon radius $R=R_0$ is constant. This is scenario \textbf{(B)}, already
discussed in \cite{ja}. Now, $M_{pr}$ decreases with density  (\ref{massbag}) by the
volume work: $M_{pr}(\varrho)=M_N- \Delta E$  at the expense of
maintaining the volume of the bag. In contrast to the discussion in \cite{ja}, the values
of s.p. energies $\varepsilon^q_N(\varrho)$ and the Fermi energies $E^q_F$ are similar in
both scenarios because the mentioned decrease of mass in scenario \textbf{(B)} is close
to the decrease of s.p. energy by the energy transfer (\ref{eq},\ref{eq1}) in
\textbf{(A)}. Therefore, both Fermi energies in \textbf{(A)} and \textbf{(B)} are smaller
than the Fermi energy  $E_F =\mu_N$ calculated for point-like nucleons (\ref{Fermi1}), by
a volume energy $p_HV_N(\varrho)$ which weakens "effectively" the repulsion between
nucleons.
\begin{figure}
\hspace{-.9cm}  \includegraphics[height=7.5cm,width=8cm]{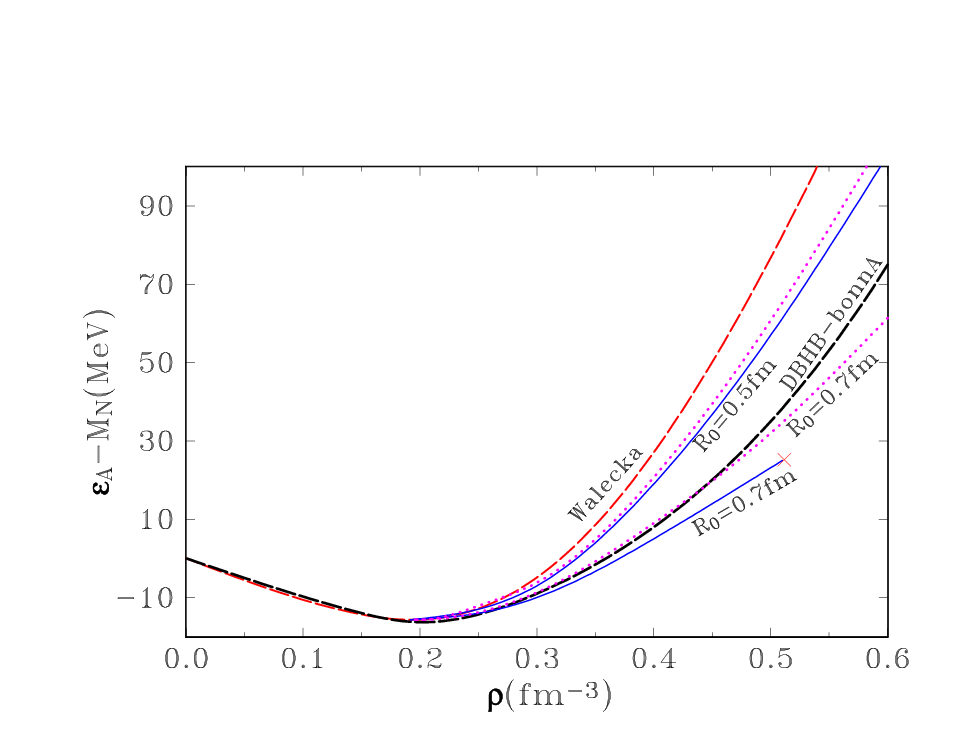} \caption{Left
panel - energy of NM above the equilibrium density for different models. Walecka
\cite{wa} and Dirac-Bruckner-Hartree-Fock (DBHF) \cite{bonn,bonn1,bonn2} calculations with the Bonn
$A$ interaction are shown as long dashes. Results for const nucleon mass (for $R=0.5;
0.7$ fm) are denoted by  dotted red lines and for const. nucleon radii (\textbf{B}) by
solid blue lines. Right: K$^{-1}$ (\ref{K}) as a function of  $R_0$ for K$^{-1}$=540
MeV.} \label{EosK}
\end{figure}
\begin{figure}
  \vspace{-10.4cm}  \hspace*{7.1cm}
\includegraphics[height=7.5cm,width=6.cm]{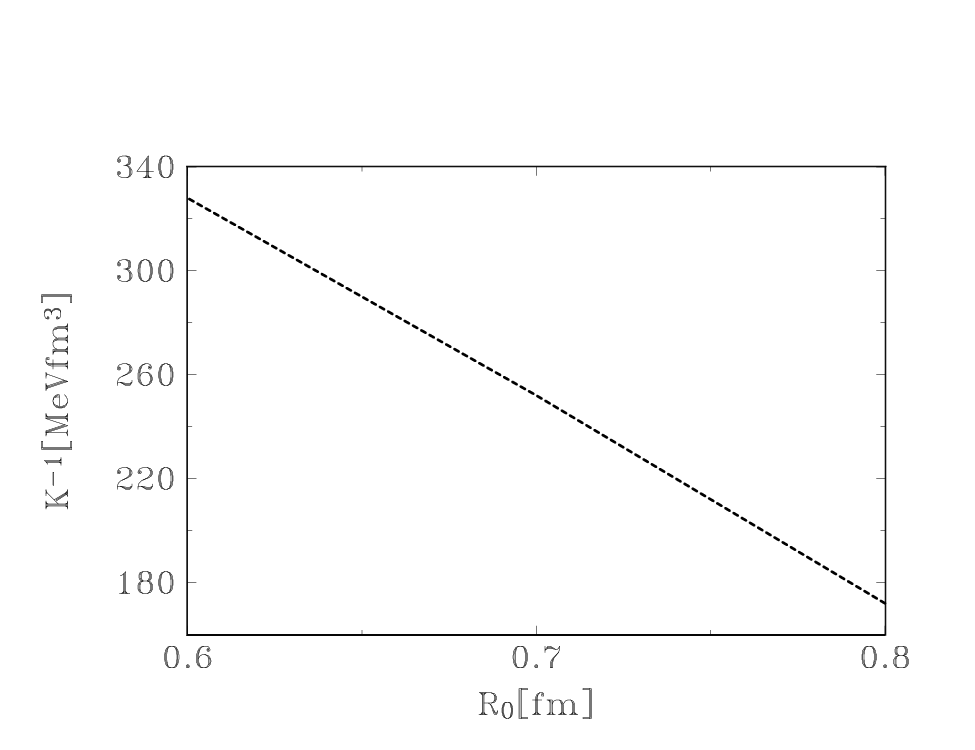}
\vspace{1.4cm}
\end{figure}
\section{Results}
 In case (\textbf{A}) we can calculate nuclear compressibility $K_q^{-1}$ using (\ref{eq}) and (\ref{eq1}):
\begin{eqnarray}
 \hspace{-.5cm}K_q^{-1}\doteq9\varrho^2\frac{\partial^2\varepsilon^q_N}{\partial\varrho^2}|_{\varrho=\varrho_0}=9\varrho^2
\frac{\partial E^q_F}{\partial\varrho} [1-\varrho
V_N(\varrho)]/\varrho|_{\varrho=\varrho_0}= K^{-1}\frac{[1-\varrho_0
V_N(\varrho_0)]}{[1+\varrho_0 V_N(\varrho_0)]}\label{K}
\end{eqnarray}
where a standard
$K^{-1}\doteq9\varrho^2\frac{\partial^2\varepsilon_N(\varrho)}{\partial\varrho^2}|_{\varrho=\varrho_0}$
has no finite volume effects.  The dependence of $K_q^{-1}$ from $R_0$
  is displayed in Fig. \ref{EosK}, right panel.
 In case (\textbf{B}) $M_{pr}=M_N-\Delta E$ and near equilibrium $M^*_{pr}\simeq M^*_N\gg P_F$,
 therefore $\sqrt{P_F^2\!+\!M^{*2}_{pr}}\simeq M^{*}_{pr}\!+\!
P_F^2/(2M^{*}_{pr})$. Thus, a decrease of $E^q_F(\varrho)$ by
$\Delta E$ (\ref{eq1}) in (\textbf{A}) corresponds in scenario (\textbf{B}) to an equivalent decrease
of the Fermi energy by smaller mass $M^*_{pr}\simeq M^*_N-\Delta E$.
 Consequently, a similar value of the compressibility (the
initial stiffness of EoS) is obtained by the formula (\ref{K}) in both scenarios and is assigned to
 a reasonable nucleon radius $\sim0.7$ fm, which strongly supports
 the  premise that volume corrections are physically responsible for nonlinear terms (\cite{boguta,boguta1}) in the scalar mean field.

\begin{figure} \vspace{-2mm}\hspace*{-0.4cm}
\includegraphics[height=9.1cm,width=13.2cm]{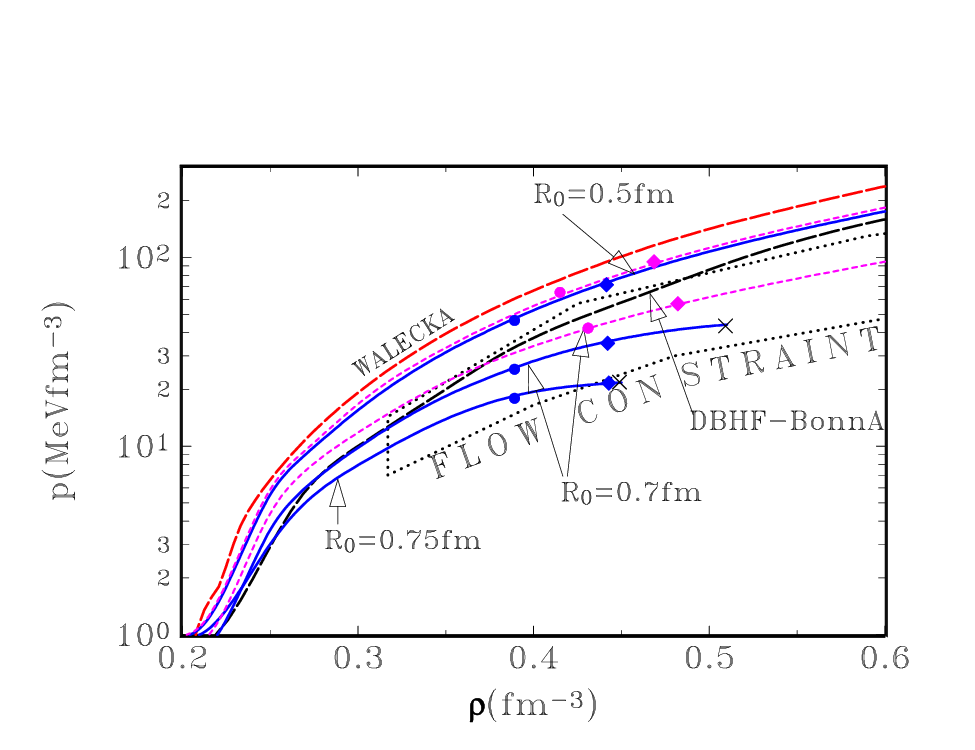} \vspace{-2mm}\caption{The
uniform pressure $p(\varrho)=\varrho^2 \varepsilon^{'}_A(\varrho)$ versus density
$\rho$. The area indicated by the ``flow constraint'' taken from \protect\cite{pawel}
determines the allowed course of the EoS, using an analysis which extracts
information from the matter flow in heavy ion collisions from the high pressure
obtained there. Walecka \cite{wa} and DBHF calculations \cite{bonn,bonn1,bonn2} with a Bonn $A$
interaction are shown as long dashes. Results for const. radii $R_0=0.5;
0.7; 0.75;$ fm are denoted by solid blue lines, results for const. mass
with $R_0=0.5; 0.7;$ fm are denoted by short red dashes. For dots, diamonds and crosses see text.}
\label{EOS}
\end{figure}

\noindent For higher densities the satisfying EoS
 is obtained for the same reasonable $R_0\sim 0.7$ fm and follows in Figs. \ref{EosK},\ref{EOS} a realistic DBHF
calculation\cite{bonn,bonn1,bonn2} in both scenarios: (\textbf{A}) (red dotted lines) and (\textbf{B}) (blue solid lines). The EOS in (\textbf{A}) is slightly stiffer because the finite volume corrections are
smaller for the decreasing nucleon volume in this case.
For $R_0<0.6$ fm the EOS is above the allowed region (Fig. \ref{EOS}) determined
by the ``flow constraint"\cite{pawel} and is below that region for $R_0>0.75$.
\noindent (Note
that in GCM \cite{bag,bag1} or QMC models \cite{Hua2,Hua,Hua1} the bag radius remains almost constant
with a reasonably stiff EOS.)

\noindent The deconfinement transition will start when the bag constant vanishes (cf. (\ref{pressure})):
\begin{eqnarray}
B(\varrho)=B(\varrho_0)(R_0/R)^4-p_H)=0,    \label{B}
\end{eqnarray}
first in the scenario (\textbf{B}) then in (\textbf{A}) where the $R(\varrho)$ decreases.
The corresponding critical densities $\varrho\simeq(0.38,0.43)$ fm$^{-3}$, obtained in case (\textbf{B}) depend from the initial values of the bag constant and are marked for two bag constants $B(\varrho_0)=(60,100)$ MeVfm$^{-3}$
  by blue dots and diamonds respectively, on solid lines in Fig. \ref{EOS}. The following self-consistent condition determines the alignment density $\varrho_{al}$ where the energy
densities outside and inside a nucleon are equal:
\begin{eqnarray}
\varrho_{al}\varepsilon^q_N(\varrho_{al})=
\!M_{pr}(\varrho_{al})/V_N(\varrho_{al}) \label{enthnuc14}
\end{eqnarray}
The condition (\ref{enthnuc14}) is fulfilled in scenario (\textbf{B}). Corresponding alignments densities depend from the nucleon radii and are marked as crosses for $R_0=(0.7,0.75)$ fm in Figs. \ref{EosK},\ref{EOS} at
the end of the solid lines.
Fig. 4 illustrates how the nuclear energy density $\varrho\varepsilon^q_N(\varrho)$ grows with density
while the nucleon energy density $M_{pr}(\varrho)/V_N(\varrho)$ in scenario (\textbf{B}) declines and finally both energy densities for $\varrho\sim0.5$ fm$^{-3}$ are equal.  For that density, nucleon bags with constant $R_0\sim0.7$  starts to overlap in case (\textbf{B}) and multi-quark bags would be possibly formed. The alignment density depends strongly on the nucleon radius, in turn the points where $B(\varrho)$=0 depend mainly from the starting value $B(\varrho_0)$ (\ref{B}).   For example, for $R_0=0.75$ fm the alignment density $\varrho_{al}=0.44$ fm$^{-3}$, shown in Figs. 4 , almost coincides in Figs. \ref{EOS} with a vanishing bag constant $B(\varrho_0)=100$ MeV fm$^{-3}$.
Therefore, scenario (\textbf{B}) with constant nucleon radius and the
gradual alignment of the energy densities inside and outside the bag suggests the crossover transition below $\varrho=0.45$ fm$^{-3}$.
\noindent However, such a transition around $\varrho\simeq 0.4$ fm$^{-3}$ is not observed in heavy ion experiments. Also in neutron stars \cite{stars,stars2}, for that density of star core we would expected for the quark core to decrease the radius of the star, but such a decrease is not expected\cite{stars1} in comparison to lighter stars with a standard neutron core.
\begin{figure}
\hspace*{1.6cm}
\includegraphics[height=6.2cm,width=9.5cm]{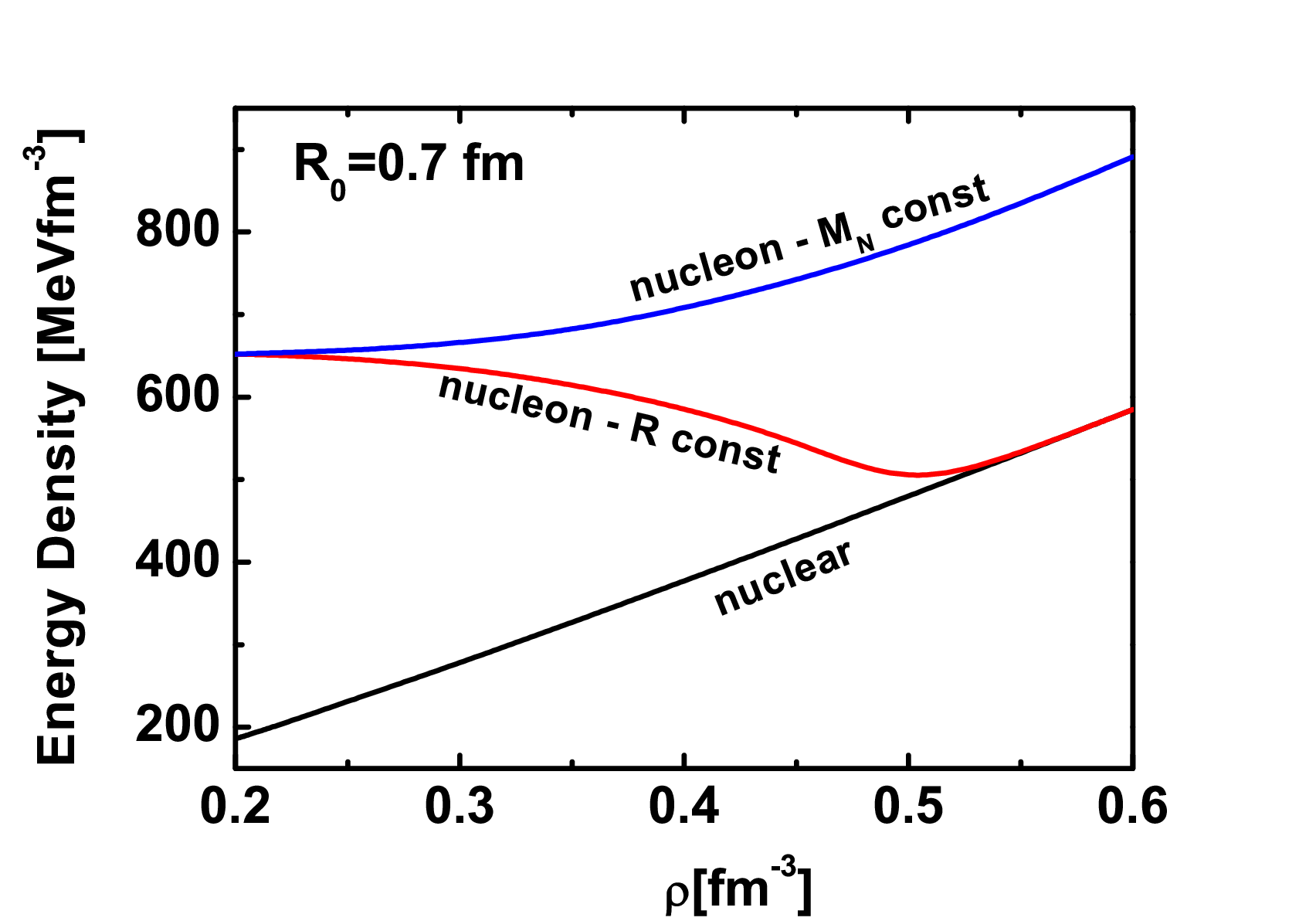}
\caption{
Energy density inside nucleons as a
function of the nuclear density for $R_0=0.7$ fm in two cases:  (A) const. nucleon mass
(red line) and (B) const. radius (blue line). The density of nuclear energy (black line)
is shown for reference.} \label{radi}
\end{figure}
Still for constant nucleon mass, scenario (\textbf{A}), a nucleon volume decreases with $\varrho$, therefore nucleon bags do not overlap for large density and the energy density of the nucleon increases due to the energy transfer into nucleon bags. The red marks (dots-diamonds) situated on the dash red lines in Fig. \ref{EOS} correspond to density range $\varrho\simeq(0.44-0.48)$ fm$^{-3}$
 where the  bag constant, starting from initial values of $B(\varrho_0)$=(60-100) MeV fm$^{-3}$, vanishes. Contrary to case (\textbf{B}),
Fig. \ref{radi} shows in case (\textbf{A}) a big difference in the energy densities outside
and inside a nucleon, which favors for $\varrho\gtrsim0.5$\hspace{1mm}fm$^{-3}$ the deconfinement as a first order phase transition from  hadron to quark matter.

Calculation of the symmetry energy (\ref{sym}) \cite{Kubis} in scenario (\textbf{A}) gives the value $E_s=24.8$ MeV for $R_0=0.7$ fm, which is a few MeV too low from phenomenological extrapolation ${E_s}^{exp}=30.53$ MeV{\cite{chin}}. In turn, a symmetry slope (\ref{sym}) parameter $L=88MeV$ is higher then the phenomenologically  extrapolated value $L^{exp}=52.5\pm 20$ MeV \cite{chin}. It is straightforward to include the additional coupling $g_{\rho}$ to the $\rho$ meson \cite{Kubis}, which contribute only to the $E_s$ of NM (\ref{sym}) and correct the energy of asymmetric neutron matter.
\begin{eqnarray}
\hspace{3mm}E_s&=&\frac{g^2_{\rho}}{8m^2_{\rho}}\varrho\!+\!\frac{P_{F}^2}{6\sqrt{P_F^2\!+\!M^{*2}_{pr}}}\hspace{2mm} \Bigg|_{\rho=\rho_0}\label{symp}
\\ \hspace{-5mm} \frac{L}{\varrho_0}\doteq3\frac{dE_s}{d\varrho}\hspace{1mm}\bigg|_{\rho=\rho_0}&=&\frac{3 g^2_{\rho}}{8m^2_{\rho}}+ \frac{ P_{F}^2}{\sqrt{P_F^2\!+\!M^{*2}_{pr}}}\left[\frac{P_{F}^{'}(\varrho)}{P_F} - \frac{P_{F}P_{F}^{'}(\varrho)+
M^{*}_{pr}M^{*'}_{pr}(\varrho)}{P_F^2\!+\!M^{*2}_{pr}}\right]_{\rho=\rho_0} \label{sym}
\end{eqnarray}
For $(g_{\rho} /m_{\rho})^2=1.34 fm^2$ we obtain (\ref{symp}) $E_s=31$ MeV. But now the slope $L=108 MeV$ is much higher then the phenomenological estimate  $L^{exp}=52.5\pm 20$ MeV.

\noindent
The slope $L$ depends strongly (\ref{sym}) on $dM^*_{pr}/d\varrho$ which at the saturation point takes a value $dM^*_{pr}/d\varrho\mid_{\rho_0} \simeq -2000$MeVfm$^3$.
Such a large value of  $-dM^*_{pr}/d\varrho$ is directly related to the constant nucleon mass with the  strong attractive scalar potential $V_S=M^*_{pr}-M_{pr} \simeq -400 MeV \varrho/\varrho_0$. When we increase  the nucleon mass  just above the saturation density by few MeV, where the nucleon is relatively "soft" and large, we will get  $dM_{pr}/d\varrho \mid_{\rho_0}>0$ and an increase  of $dM^*_{pr}/d\varrho \mid_{\rho_0}$. Let us estimate this effect by the additional energy transfer $\Delta M= C_m p_H$ which increases to $0$ the derivative of a new effective mass $dM^*_{new}/d\varrho|_{\rho=\rho_0}=0$:
\begin{eqnarray}
\hspace{-1mm}dM^*_{new}/d\varrho|_{\rho=\rho_0}\simeq dM^*_{pr}/d\varrho+C_m dp_H/d\varrho= -2000 MeVfm^3+\frac {C_m K^{-1}_q}{9(1-\varrho V_N(\varrho))}\Bigg|_{\rho=\rho_0} \label{cm}
\end{eqnarray}
Thus for $K_q^{-1}=(200-300)MeV$, the additional transfer $\Delta M=C_mp_H$ near the saturation density (with $C_m\simeq (40-50)$ $fm^3$ (\ref{cm})) will reset $dM^*_{new}/d\varrho=0|_{\rho=\rho_0}$, which reduces the slope $L$ (\ref{sym}) from $L=108$ MeV to the accepted value $L=61$ MeV ($L^{exp}=52.5\pm 20$ MeV \cite{chin}). But the energies (\ref{eq}-\ref{eq1}) and compressibility will remain unchanged because the energy transfer $\Delta M$, decreasing the s.p. energy (like $\Delta E$), increases the nucleon mass by the same amount and the net result in the s.p. energy (\ref{eq}) is negligible.
 Such a transfer, increasing nucleon mass, decreases $R$ (\ref{massbag}). It makes the EoS a little bit stiffer (\ref{eq2}) but the decrease ($\sim1\%$ of $R$) is negligible.

 In scenario \textbf{B} the nucleon mass is decreasing in density, therefore it is difficult to fit the good value for the slope $L$ of the symmetry energy.
It is another evidence that the scenario \textbf{B} is unrealistic.
\section{Conclusions}
The presented energy transfer from the repulsive vector field to the nucleon bags ensures the nondecreasing nucleon mass with decreasing radius $R(\varrho)$  and shifts the de-confinement phase transition to the higher nuclear densities.
 The presented scenario  (\textbf{A}) with constant nucleon mass is more realistic then scenario (\textbf{B}) \cite{ja}  without energy transfer \footnote{In \cite{ja} we did not consider the energy transfer, therefore we obtain the softer EoS and the good value of $K_q^{-1}$ only in scenario (\textbf{B}). Here a significant development of the previous work has been presented.} which predicts an unobserved crossover transition for $\varrho\simeq0.4$ fm$^{-3}$. The energy transfer, equal to the volume energy $\Delta E=p_HV_N$, provides the constant nucleon mass, the good values of the compressibility $K_q^{-1}\sim (250-350)$ MeV \cite{K,K1} and the symmetry energy $E_s=31$ MeV. The additional energy transfer at the saturation region above the equilibrium density, which increases slightly the nucleon mass , reduces the value of  the slope of the symmetry energy from $L\simeq 108$ MeV to $L\simeq 61$ MeV \cite{chin}.
The presented model of NM determines the changes of nucleon properties like its mass and radius, fitted to the saturation properties of NM. Thus it is the alternative to the widely exploit RMF models with the rich virtual meson structure which contains nonlinear terms \cite{menezes,boguta} in the scalar potential and density dependent couplings to mesons, exchanged between point like nucleons. For higher density and reasonable nuclear radii $R_0\sim0.7$ fm, the presented
volume corrections convert the unrealistic, very stiff EOS of the scalar-vector model \cite{wa} to a suitable EOS (Figs. \ref{EosK},\ref{EOS}), which follows realistic DBHF calculations \cite{bonn,bonn1,bonn2,blaszka,blaszka1}.
The presented model contains the nucleons as extended objects in the mean field. We show that the changes of the physical parameters, like the nucleon radius or the mass, are very sensitive to  experimental constrains like nuclear compressibility, symmetry energy and its slope.
It will be interesting to include the decreasing  nucleon radius in other calculations (e.q. ab-initio) of equation of state for a dense nuclear and neutron matter.

\section*{Acknowledgements}
\noindent We thank the referee for suggesting valuable improvements of our manuscript.

\end{document}